\documentclass[reprint,showpacs,showkeys,preprintnumbers,amsmath,amssymb,jap,aip,floatfix,groupedaddress]{revtex4-1}
\usepackage[utf8]{inputenc}
\usepackage[dvips]{graphicx}
\usepackage[dvips]{graphics}
\usepackage{dcolumn}
\usepackage[mathlines,pagewise]{lineno}
\usepackage{amsmath}
\usepackage{placeins}
\usepackage[breaklinks=true,colorlinks=true,linkcolor=blue,urlcolor=blue,citecolor=blue]{hyperref}

\begin{document}
%
\title{Anisotropic exchange in GdGa}\label{title}
\author{V. S. R. de Sousa}
\email{vsousa@uerj.br}
\author{E. P. Nóbrega}
\author{P. O. Ribeiro}
\author{B. P. Alho}
\author{P. J. von Ranke}
\affiliation{%
Instituto de F\'isica Armando Dias Tavares, Universidade do Estado do Rio de Janeiro (UERJ), 20550-013, Rio de Janeiro - RJ, Brazil. \\
}%
\date{\today}
%
\begin{abstract}
In this work we discuss the non-collinear ground-state reported for GdGa on the basis of a model Hamiltonian considering anisotropic exchange interactions. We show that a competition between intra and inter-sublattice exchange interactions can lead to a canted structure, which competes with a collinear (ferro or antiferromagnetic) order. The mean-field thermodynamic analysis of the model for $S=7/2$ shows a good agreement between calculated  and reported experimental data of the canting angle and isothermal entropy change for GdGa.
\end{abstract}
\maketitle
%

\section{Introduction}
	GdGa crystallizes in the orthorhombic CrB-type structure, space group Cmcm ($\#$63), in which Gd and Ga occupy 4c site\cite{baezinger,buschow,zhang,susilo}. It orders ferromagnetically around 200 K and undergoes a spin-reorientation transition at a lower temperature. The ferromagnetic order is collinear between T$_\text{C}$ and T$_\text{SR}$, and for temperatures below T$_{SR}$ it becomes non-collinear, with Gd 4c site being split into two non-equivalent ones. This spin reorientation has been studied experimentally with both M\"ossbauer spectroscopy and neutron powder diffraction\cite{delyagin,susilo}.


Since Gd$^{3+}$ have a spherical symmetrical charge distribution, resulting from the half-filled 4f-shell configuration, we can rule out the influence of the crystal field interaction on the spin reorientation in GdGa. One may speculate about the influence of the classical dipole-dipole interaction\cite{luttinger,johnston}, which leads to anisotropic effects, but in the present case the reported reorientation temperatures are too high to be explained only by this interaction. The antisymmetric exchange\cite{dzy,moriya}, which tends to orient two interacting spins perpendicularly, destabilizing collinear ferromagnetic or antiferromagnetic structures, and can cause a canting of the spins in different sites\cite{blugel}, may be ruled out on a symmetry basis: since the middle of the bonds between Gd-atoms in GdGa are inversion centers the Dzyaloshinskii–Moriya interaction is null\cite{lacroix}. As pointed out by Delyagin et al.\cite{delyagin}, this reorientation might be a consequence of anisotropic exchange interactions, which may be related with an anisotropic Gd-Ga hybridization mediating the indirect Gd-Gd interaction.

Here we report theoretical results that accounts for the major features reported from experiments in GdGa based on a microscopic model Hamiltonian that considers anisotropic exchange interactions between Gd ions. The model Hamiltonian is a two-sublattice anisotropic Heisenberg model in the presence of a magnetic field. In order to apply the model to GdGa we consider its magnetic structure as composed of alternating ferromagnetic planes as speculated by Leithe-Jasper and Hiebl\cite{jasper}. The exchange parameters were obtained semi-empirically following a detailed analysis of the system behavior near the phase transitions. The calculated spin canting angle and isothermal entropy change show a good agreement with the reported experimental data for GdGa.
%

\section{Formalism}\label{sec:formalism}

\begin{figure}[htb!]
\centering
\includegraphics[width=0.75\linewidth]{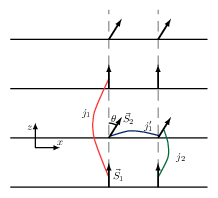}
\caption{\label{fig:sketch} Sketch of the layered magnetic structure considered for the model Hamiltonian. }
\end{figure}

We consider the following anisotropic Heisenberg model in the presence of a magnetic field:
\begin{equation}
	{\cal H} = -\sum_{ij}\left({ j_{ij}^\perp S_i^zS_j^z + j_{ij}^{\parallel}\left(S_i^xS_j^x+S_i^yS_j^y\right)}\right) - \vec B_0\cdot \sum_{i}{g_i\mu_B\vec S_i}.\label{eq:heisenberg}
\end{equation}
where $\bar{j}_{ij}^\alpha$ ($\alpha=\perp,\parallel$) are the exchange parameters, $\vec B_0=B_0\hat z$ is the applied magnetic field, $g_i$ the Land\'e factor of the i-th spin and $\mu_B$ the Bohr magneton. We also assume a magnetic structure composed of alternanting ferromagnetic layers in which the spins are either parallel to a global z-axis or canted from this axis by an angle $\theta$. This arrangement  is sketched in Fig.~\ref{fig:sketch}.

For the sake of simplicity, the exchange interaction between spins in both sublattices are taken as: $j_{11}^\parallel=j_1$, $j_{11}^\perp=0$; $j_{22}^\parallel=0$, $j_{22}^\perp=j'_1$; $j_{12}^\parallel=j_2$, $j_{12}^\perp=0$. This implies that the intra-layer exchange between spins in the layer composed of spins parallel to $z$ is $j_1$; the intra-layer exchange between canted spins is along an axis perpendicular to $z$; and the inter-layer exchange $j_2$ is along $z$. For convenience, we name the spins in the layer composed of spins parallel to $z$ as $\vec S_1$, and the canted ones as $\vec S_2$, and we also limit the calculations to the $XZ$ plane.

Considering the above set of exchange parameters, we may decomposed relation~(\ref{eq:heisenberg}) in two-sublattice Hamiltonians, i.e., ${\cal H}={\cal H}_1+{\cal H}_2$, so that

\begin{subequations}\label{eq:subs}
\begin{equation}
{\cal H}_1 = -j_1\sum_{\langle ij\rangle}{S^z_{1i} S^z_{1j}} - j_2\sum_{\langle ij\rangle}{S^z_{1i} S^z_{2j}} - \mu_B B_0\sum_i{g_{1i}S^z_{1i}},
\end{equation}
\begin{equation}
	{\cal H}_2 = -j'_1\sum_{\langle ij\rangle}{S^x_{2i} S^x_{2j}} - j_2\sum_{\langle ij\rangle}{S^z_{2i} S^z_{1j}} - \mu_B B_0\sum_i{g_{2i}S^z_{2i}}.
\end{equation}
\end{subequations} 

In the above relations the summation is restricted to nearest-neighbors (NN). Note that in the present form, relations~(\ref{eq:subs}) could be used to study a layered structure of unequal moments. Here we make the further assumption that $S_1$=$S_2$=$S$ ($g_1$=$g_2$=$g$) and $j'_1$=$j_1$, this last one is justifiable since we are considering a magnetic structure of equal magnetic moments.
%
\subsection{Classical ground state}
In a classical picture, the spin operators in both sublattices will be given by: $\vec S_1$ = $S_1^z \hat z$ = $S\hat z$; $\vec S_2$ = $S_2^x\hat x + S_2^z\hat z$, $S_2^x$ = $S\sin\theta$ and $S_2^z$ = $S\cos\theta$. The total energy $E$ = $E_1+E_2$ is therefore
\begin{equation}\label{eq:classen}
E=-j_1S^2(1+\sin^2\theta) - 2j_2S^2\cos\theta - g\mu_BB_0S(1+\cos\theta).
\end{equation}

By minimizing this energy with relation to $\theta$ one gets the following critical angles in the absence of an external magnetic field: $\theta=0$, $\theta=\pi$ and $\theta=\cos^{-1}\left(j_2/j_1\right)$. Note that this gives us two collinear solutions, corresponding to ferromagnetism ($\theta=0$) and antiferromagnetism ($\theta=\pi$); and a non-collinear solution that depends on the ratio $j_2/j_1$. For the case in which both $j_2$ and $j_1$ are positive the antiferromagnetic (AFM) solution is always unstable, whereas there will be a competition between ferromagnetic (FM) and non-collinear magnetic (NCM) states since: (i) $j_2 > j_1$ favors FM and $j_2<j_1$ favors a canted (non-collinear) magnetic structuce. Conversely, when $j_2$ is negative the FM phase is always unstable and for the case $j_1 > 0$ one may observe a competition between AFM (favored if $|j_2| > j_1$) and NCM (favored if $|j_2| < j_1$) states.

\begin{figure}[htb!]
\centering
\includegraphics[width=0.75\linewidth]{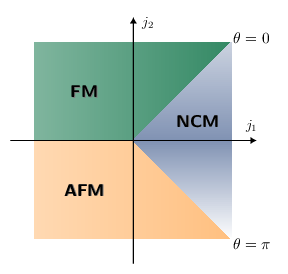}
\caption{\label{fig:classdiag} Classical ground state magnetic phase diagram. The competition between the different magnetic structures is depicted as a function of $j_1$ and $j_2$.}
\end{figure}

The classical ground state phase diagram is depicted in Fig.~\ref{fig:classdiag}. The transition between the different magnetic states is marked by the reversible lines $j_2=j_1$ (FM $\rightleftarrows$ NCM), $j_2=-j_1$ (AFM $\rightleftarrows$ NCM) and $j_2=0$ (FM $\rightleftarrows$ AFM, with $j_1<0)$).

There is a critical magnetic field ($B_c$), applied along $z$, for which the FM state becomes more favorable than the NCM one, even if $j_1>j_2$. One can show from~(\ref{eq:classen}) that this field will be given by
\begin{equation}\label{eq:bcritic}
B_{c} = \frac{2(j_2-j_1)}{g\mu_B}S.
\end{equation}

This classical picture illustrates that an anisotropic exchange interaction may be responsible for the non-collinear magnetic state observed in GdGa. 

\subsection{Mean field analysis for $S=1/2$}\label{subsec:mfield}

Here we discuss the above model for an ensemble of spins with $S=1/2$ in the mean field approximation. This allows us to obtain equations of state that, by its turn, can be evaluated at some critical conditions in order to stablish possible phase transitions.

In the mean-field approximation, equations (2) are rewritten in the form of an effective field acting on each sublattice, so that, ${\cal H}_1=-b_{1z}S_1^z$ and ${\cal H}_2=-b_{2z}S_2^z-b_{2x}S_2^x$, where $b_{1z}=b_0+2z_1j_1\langle S_1^z\rangle+2z_2j_2\langle S_2^z\rangle$, $b_{2z}=b_0+2z_2j_2\langle S_1^z\rangle$, $b_{2x}=2z_1j_1\langle S_2^x\rangle$, $b_0=g\mu_B B_0$. The eigenvalues of ${\cal H}_1$ and ${\cal H}_2$ will be given by $\varepsilon_1^{(1)}=-Sb_{1z}$, $\varepsilon_2^{(1)}=+Sb_{1z}$, $\varepsilon_1^{(2)}=-S\sqrt{b^2_{2x}+b^2_{2z}}$, $\varepsilon_2^{(2)}=+S\sqrt{b^2_{2x}+b^2_{2z}}$.

Once we have obtained the eigenvalues, we can proceed by calculating the mean thermodynamic values of the operators $S_1^z$, $S_2^x$ and $S_2^z$ using the relation:
	\begin{equation}
 		\langle S_l^k\rangle = \frac{1}{Z_l} \sum_l{ \left(-\frac{\partial \varepsilon_i^{(l)}}{\partial b_{lk}}\right) e^{-\beta\varepsilon_i^{(l)}} },
 	\end{equation}
where $l$=1,2, $k$=$x$,$z$, $Z_l=\sum_i{e^{-\beta\varepsilon_i{(l)}}}$ is the canonical partition function, $\beta=1/k_BT$ ($k_B$ is the Boltzmann constant and T the temperature). After some algebra one gets the following results
\begin{subequations}\label{eq:eosmag}
\begin{equation}
 \langle S_1^z\rangle = S\tanh{(\beta Sb_{1z})},
 \end{equation}
\begin{equation}
 \langle S_2^x\rangle = S\frac{b_{2x}}{\sqrt{b^2_{2x}+b^2_{2z}}}\tanh{\left(\beta S\sqrt{b^2_{2x}+b^2_{2z}}\right)},
 \end{equation}
\begin{equation}
 \langle S_2^z\rangle = S\frac{b_{2z}}{\sqrt{b^2_{2x}+b^2_{2z}}}\tanh{\left(\beta S\sqrt{b^2_{2x}+b^2_{2z}}\right)}.
 \end{equation}
\end{subequations}

 Relations~(\ref{eq:eosmag}) form a group of magnetic equations of state that one can use to calculate the dependence of the magnetization as a function of temperature and magnetic field. Here we are mainly interested in the behavior around the spin reorientation temperature ($T_{SR}$) and the Curie temperature ($T_C$). 

$T_{SR}$ marks the transition from the noncollinear to a collinear magnetic state (FM if $j_2>0$ or AFM if $j_2<0$). At this temperature both sublattices magnetization becomes parallel to the z-axis, which implies that  $\langle S_2^x\rangle \rightarrow 0$ at $T_{SR}$. Using this condition in~(\ref{eq:eosmag}b) one finds that (for $B_0=0$) the reorientation temperature have the following dependence on the exchange parameters
\begin{equation}\label{eq:tsr}
T_{SR} = \frac{4\sigma_1S^2}{\ln\left(\frac{1+a}{1-a}\right)} \frac{z_2j_2}{k_B},
\end{equation}
where $a=\sigma_1z_2j_2/z_1j_1$ and $\sigma_1=\langle S_{1z}\rangle$ is the expectation value of S$_{1z}$ evaluated from Eq.~(\ref{eq:eosmag}a) at T$_\text{SR}$. This implies that equation~(\ref{eq:tsr}) is calculated self-consistently, in a similar fashion as the critical temperature of the Oguchi method\cite{oguchi,smart}. One may obtain the Curie (or N\'eel) temperature from relations~$(\ref{eq:eosmag}a)$ and~(\ref{eq:eosmag}c) in the limit of small arguments, so that
\begin{equation}\label{eq:tcurie}
T_{C(N)} = \frac{4S^2a}{(-1+\sqrt{1+4a^2})}\frac{z_2j_2}{k_B}.
\end{equation}

One may also consider the limit $T\rightarrow 0$. In this case, $\langle S_{1z}\rangle=S$, and we can define the ratio between $\langle S_2^x\rangle$ and $\langle S_2^z\rangle$ as the tangent of the canted angle. After some algebra, one gets
\begin{equation}\label{eq:tanq}
\tan\theta = \sqrt{\left(\frac{z_1/z_2}{j_2/j_1}\right)^2 -1}.
\end{equation}

\begin{figure}[htb!]
\centering
\includegraphics[width=0.75\linewidth]{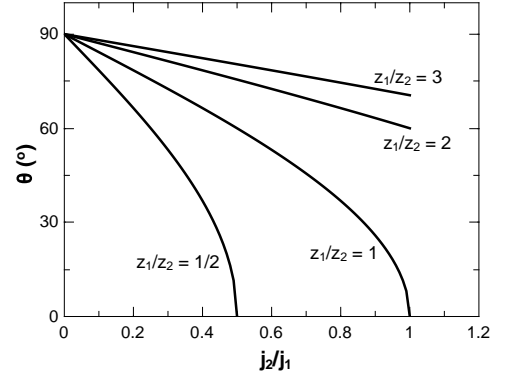}
\caption{\label{fig:angles} Canting angle as a function of the ratio $j_2/j_1$ for several coordinations $z_1/z_2$ = 2 (simple cubic and simple tetragonal lattices), 1/2 (face-centered and body-centered cubic lattices), 3 (hexagonal lattice). $\theta=90^\circ$ implies non-interacting sublattices and $\theta=0$ ferromagnetism.}
\end{figure}

In the above relation the canting angle depends on the coordination number as shown in Fig~\ref{fig:angles}.

\subsection{Critical temperatures for $S>1/2$}\label{app:temp}
In order to obtain the spin reorientation temperature for any $S$-value, we consider the Hamiltonian
\begin{equation}
{\cal H} = -b_{2z}S_2^z - b_{2x}S_2^x,
\end{equation}
as composed of an unperturbed
\begin{equation}
{\cal H}_0 = -b_{2z}S_2^z
\end{equation}
and a perturbed
\begin{equation}
{\cal H}' = -b_{2x}S_2^x = -\frac{b_{2x}}{2}(S^++S^-)
\end{equation}
contribution. This is well justified since close to $T_\text{SR}$ the perpendicular magnetization (and, henceforth, the effective field along $x$) approaches zero. The unperturbed eigenenergies $\varepsilon_m^{(0)}=-b_{2z}m$ are easily obtained, with the corresponding eigenvectors $|\varepsilon_m^{(0)}\rangle=|S,m\rangle$ ($m=-S,\dots,S$). The first order correction to the eigenenergies due to the perturbation $\varepsilon_m^{(1)}=\langle\varepsilon_m^{(0)}| {\cal H}_1 |\varepsilon_m^{(0)}\rangle$ is null, and to second order the correction will be given by
\begin{equation}
\varepsilon_m^{(2)}=-\frac{b_{2x}^2}{2b_{2z}}m.
\end{equation}

The mean thermodynamic value of $S_{2x}$ can then be obtained from relation
\begin{equation}
\langle S_{2x}\rangle = \frac{1}{Z^{(0)}}\sum_{m=-S}^S {\left(-\frac{\partial\varepsilon_m^{(2)}}{\partial b_{2x}}\right)e^{-\beta\varepsilon_m^{(0)}}},
\end{equation}
where $Z^{(0)}=\sum_{m=-S}^S{e^{-\beta\varepsilon_m^{(0)}}}$. One can show that the above equation can be cast in the form
\begin{equation}
\langle S_{2x}\rangle = S\frac{b_{2x}}{b_{2z}}B_S\left(\beta Sb_{2z}\right),
\end{equation}
where $B_S(x)$ is the Brillouin function. This result is valid close to the spin reorientation temperature. We can therefore substitute $b_{2x}$ and $b_{2z}$ above and consider $T=T_\text{SR}$ to get the following equation
\begin{equation}
T_\text{SR} = \frac{2\sigma_1S^2}{B_S^{-1}(a)}\frac{z_2j_2}{k_B}.
\end{equation}

We may now consider the following expansion for the inverse Brillouin function\cite{kroger}
\begin{equation}
B_S^{-1}(a)=\frac{\kappa(a)}{2}\ln\left(\frac{1+a}{1-a}\right),
\end{equation}
where
\begin{equation}
\kappa(a)=\frac{15 -11(1-\epsilon)(1+2\epsilon)a^2}{5+10\epsilon-(1-\epsilon)[5+11\epsilon(1+2\epsilon)]a^2}
\end{equation}
with $\epsilon=1/2S$. Therefore the spin reorientation temperature for any $S$ value will be given by
\begin{equation}\label{eq:tsrapp}
T_{SR} = \frac{4\sigma_1S^2}{\kappa(a)\ln\left(\frac{1+a}{1-a}\right)}\frac{z_2j_2}{k_B},
\end{equation}
which differs from~({\ref{eq:tsr}}) by a factor $\kappa(a)$ in the denominator. Note that, if $\epsilon=1 \rightarrow \kappa(a)=1$, and we recover the result obtained for $S=1/2$.

One may also show that the equation for $T_\text{C(N)}$ is generalized by replacing $S^2$ by $S(S+1)/3$ in~(\ref{eq:tcurie}), therefore
\begin{equation}\label{eq:tcurieapp}
T_{C(N)} = \frac{4S(S+1)}{3k_B}\frac{a}{(-1+\sqrt{1+4a^2})}z_2j_2.
\end{equation}

Relations~(\ref{eq:tsr}) and~(\ref{eq:tcurie}), or~(\ref{eq:tsrapp}) and~(\ref{eq:tcurieapp}), allows one to obtain for some experimentally determined spin reorientation and critical temperatures the values of $j_2$ and $j_1$. These parameters can be used as input values in the model and one may evaluate the equations of state for the magnetization, and also obtain other thermodynamic quantities (e.g., specific heat) as a function of temperature and magnetic field.

With these semi-empirical exchange parameters one can numerically estimate the canting angle using
\begin{equation}\label{eq:theta}
\theta = \tan^{-1}\left(\frac{\langle S_{2x}\rangle}{\langle S_{2z}\rangle}\right)
\end{equation}

 We point out that the critical magnetic field, at which the NCM state becomes less stable than the FM one, will also be given by~(\ref{eq:bcritic}) with a slight modification $j_1\rightarrow z_1 j_1$ and $j_2\rightarrow z_2j_2$.
%
%
\section{Results and Discussions}

	Here we consider the model described in~\ref{sec:formalism} to study the spin reorientation transition and the thermomagnetic behavior of the compounds GdGa, which shows a non-collinear to ferromagnetic phase transition. We take the Land\'e factor and spin angular momentum given by Hund's rules: $g=2$ and $S=7/2$. From the reported crystallographic structure of GdGa, gadolinium atoms have two first neighbors and four second neighbors, therefore $z_1=2$ and $z_2=4$.

	We have calculated the isothermal entropy change by the standard procedure described, for instance, in Ref. [\onlinecite{nilson2010}]. The magnetic entropy of each sublattice is obtained from the relation $S_\text{mag} = Nk_B\left( \ln Z_l + \beta E_l\right)$, where $E_l = \langle \mathcal{H}_l\rangle$ represents the mean thermodynamic value of $l$-th Hamiltonian in (\ref{eq:subs}). In order to compare the calculations of the isothermal entropy change with experimental data we adopt a procedure in which the applied magnetic field direction is varyied over a unit sphere, in an averaging procedure used to simulate a polycrystalline sample\cite{alho2012}.
	
	Table~\ref{tab:example} lists some reported experimental critical temperatures for GdGa as well as the experimental relative canting angle between Gd atoms in the non-collinear phase ($\theta_\text{exp}$). We also list the corresponding exchange parameters obtained from equations~(\ref{eq:tsrapp}) and~(\ref{eq:tcurieapp}) and the canting angle calculated from~(\ref{eq:theta}). One can note a good agreement between calculated and experimental canting angles, showing that in fact the mechanism responsible for the non-collinear magnetic state in GdGa is an anisotropic exchange.
\begin{table}[h]
\begin{ruledtabular}
\caption{\label{tab:example}Experimental Curie temperature (T$_C$), spin reorientation temperature (T$_{SR}$) and relative angle between Gd moments ($\theta_\textrm{exp}$), and calculated intra-sublattice exchange parameter ($j_1$), inter-sublattices exchange parameter ($j_2$) and canting angle ($\theta_\text{calc}$).}
\begin{tabular}{c c c c c c c}
T$_C$ & T$_{SR}$ & $\theta_\textrm{exp}$ ($^\circ$) & Ref. &$j_1$ (meV) & $j_2$ (meV) & $\theta_\text{calc}$($^\circ$) \\\hline
190 & 68 & 38\footnote{From neutron powder diffraction data.} & [\onlinecite{susilo}] & 0.514 & 0.226 & 28 \\
183 & 100 & 45\footnote{From $^{119}$Sn M\"ossbauer spectroscopy data.} & [\onlinecite{delyagin}] & 0.533 & 0.199 & 42 \\
\end{tabular}
\end{ruledtabular}
\end{table}

\begin{figure}[htb!]
\centering
\includegraphics[width=0.9\linewidth]{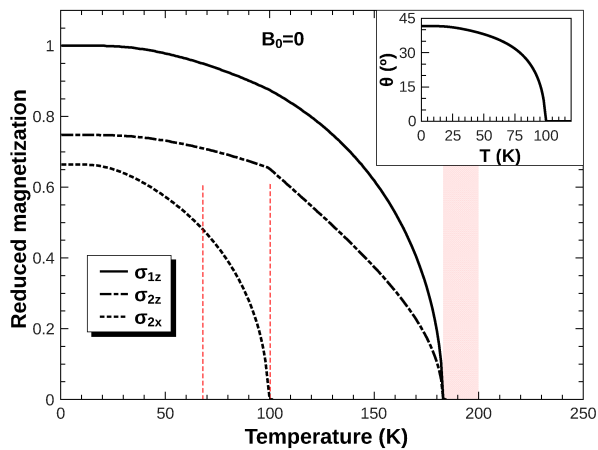}
\caption{\label{fig:magn} Zero field reduced magnetization sublattices as a function of temprature calculated using $j_1=0.533$ meV and $j_2=0.199$ meV00. The vertical dashed lines mark the reported experimental T$_\text{SR}$ and the shaded area the range of reported experimental Curie temperatures for GdGa. The inset shows the evolution of the canting angle as a function of temperature.}
\end{figure}

	Figure~\ref{fig:magn} shows the calculated zero-field reduced sublattices magnetization ($\sigma_k$=$\langle S_k\rangle/S$) as a function of temperature using the exchage parameters $j_1=0.533$ meV and $j_2=0.199$ meV. At very low temperatures, $\sigma_{1z}$ is fully satured and the components of the canted sublattice have the values $\sigma_{2z}=\cos\theta_\text{calc}$ and $\sigma_{2x}=\sin\theta_\text{calc}$. As temperature increases the magnetization of each sublattice decreases, with the transverse magnetization (parallel to $x$) going to zero at T$_\text{SR}$. Above the reorientation temperature, the system evolves as two coupled ferromagnetic sublattices with different magnetic moments. The behaviour of the curves close to the critical temperatures is unaffected by short-range correlations since we are considering a mean-field approximation, therefore the order parameter goes to zero above these temperatures.

\begin{figure}[htb!]
\centering
\includegraphics[width=0.95\linewidth]{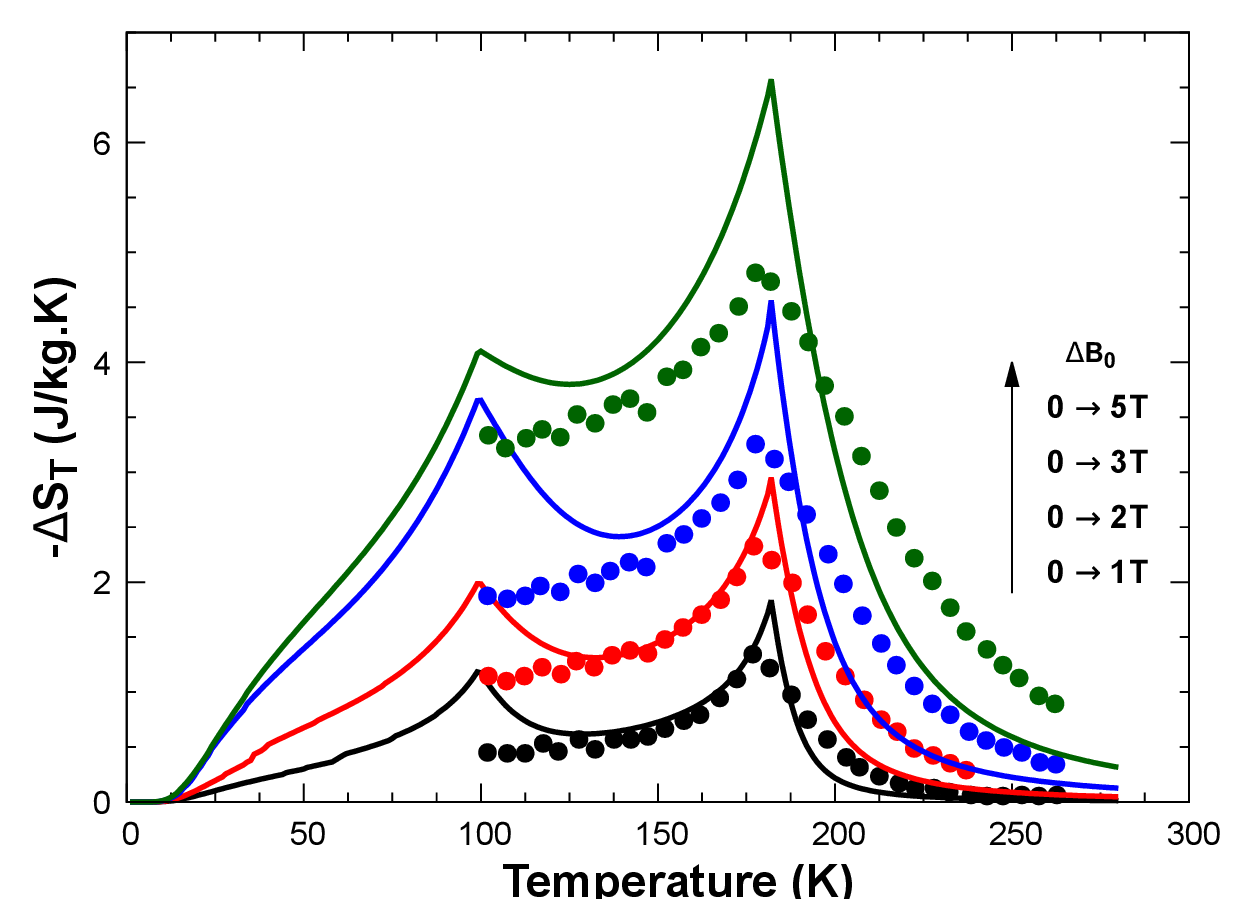}
\caption{\label{fig:gdga1} GdGa isothermal entropy change as a function of temperature for several magnetic field changes. Full circles and solid lines represent, respectively, experimental data and calculations.}
\end{figure}

	The inset in Fig.~\ref{fig:magn} shows the variation of the canting angle with temperature. At T$_\text{SR}$ both sublattices become parallel to the $z$ axis and, therefore, $\theta=0$ above this temperature. The smooth change of the canting angle as a function of temperature shows that the spin reotientation transition is of second-order, which is in accordance with an Arrott plot analysis\cite{arrott,yeung} performed by Zhang et. al\cite{zhang}, which also studied the magnetocaloric properties of GdGa, reporting a maximum value of 4.81 J/kg.K for the isothermal entropy change under a magnetic field change from 0 to 5 T. The calculated entropy change shows a good agreement with the experimental data as can be seen in Fig.~\ref{fig:gdga1}. The secondary peak presented in the theoretical curves are due to the spin reorientation transition, which has a similar impact on the isothermal entropy change as observed in the compounds Ho$_2$In\cite{zhang2009}, ErGa\cite{chen}, TbZn\cite{desousa2010}, HoZn\cite{desousa2011,lingwei}, i.e., a large relative cooling power (RCP) over a wide temperature range\cite{zhang}. The main difference being that in these compounds the spin reorientation appears as a competition between crystal field and exchange interactions, meanwhile in GdGa it is due only to the exchange mechanism.
	
\section{Summary and Outlook}

We have shown that the spin reorientation transition and the large refrigeration capacity reported for GdGa may be explained by an anisotropic two-coupled sublattices Heisenberg-like model Hamiltonian. Our mean-field results are in good agreement with the experimental data for both canting angles and isothermal entropy change. The canting angle is dependent on the ratio between intra and intersublattice exchanges, and the anisotropic exchange driven spin reorientation transition is responsible for the large observed RCP reported for GdGa\cite{zhang}. There is a previous theoretical study that reports density functional theory calculated exchange interactions\cite{katsnelson} for GdGa\cite{liu}, which cannot be directly compared to the exchange parameters adopted here because the authors have considered only a simple collinear ferromagnetic structure in their study, not considering the influence of the spin-orbit interaction in their calculations. Therefore, it would be interesting if further investigations were performed in order to describe the non-collinear magnetic ground state of GdGa from first-principles, in order to clarify the possible anisotropic hybridization between Gd and Ga atoms in this compound.

\acknowledgments{The authors thank Brazilian agencies FAPERJ and CNPq for financial support. This  study  was  financed  in  part  by  the  Coordenação  de Aperfeiçoamento  de  Pessoal  de  Nível  Superior - Brasil (CAPES) - Finance  Code  001}.%
\FloatBarrier

\bibliography{gdga_refs}
\bibliographystyle{aipnum4-2}

\appendix

\end{document}